\documentstyle[12pt]{article} \hoffset -0.5in \textwidth 6.5in \textheight
9.00in  \parskip 7pt \openup1.5\jot \parindent=0.5in
\newskip\humongous \humongous=0pt plus 1000pt minus 1000pt
\def\caja{\mathsurround=0pt}
\def\eqalign#1{\,\vcenter{\openup1\jot \caja
        \ialign{\strut \hfil$\displaystyle{##}$&$
        \displaystyle{{}##}$\hfil\crcr#1\crcr}}\,}
\newif\ifdtup

\def\eqright #1\cr{\noalign{\hfill$\displaystyle{{}#1}$}}
\def\eqleft #1\cr{\noalign{\noindent$\displaystyle{{}#1}$\hfill}}

\def\oldreffmt#1{\rlap{[#1]} \hbox to 2\parindent{}}

\def\figfmt#1{\rlap{Figure {#1}} \hbox to 1in{}}

\def\VEV#1{\left\langle #1\right\rangle}

\def\sectioneq{\def\theequation{\thesection.\arabic{equation}}{\let
\holdsection=\section\def\section{\setcounter{equation}{0}\holdsection}}}%

\newcounter{holdequation}

\def\auto{\eqno(\refstepcounter{equation}\theequation)}
\def\begineq #1\endeq{$$ \refstepcounter{equation}\eqalign{#1}\eqno
        (\theequation) $$}
\def\contlimit{\,{\hbox{$\longrightarrow$}\kern-1.8em\lower1ex
\hbox{${\scriptstyle (a\rightarrow0)}$}}\,}
\def\centeron#1#2{{\setbox0=\hbox{#1}\setbox1=\hbox{#2}\ifdim
\wd1>\wd0\kern.5\wd1\kern-.5\wd0\fi
\copy0\kern-.5\wd0\kern-.5\wd1\copy1\ifdim\wd0>\wd1
\kern.5\wd0\kern-.5\wd1\fi}}
\def\centerover#1#2{\centeron{#1}{\setbox0=\hbox{#1}\setbox
1=\hbox{#2}\raise\ht0\hbox{\raise\dp1\hbox{\copy1}}}}
\def\centerunder#1#2{\centeron{#1}{\setbox0=\hbox{#1}\setbox
1=\hbox{#2}\lower\dp0\hbox{\lower\ht1\hbox{\copy1}}}}
\def\lsim{\;\centeron{\raise.35ex\hbox{$<$}}{\lower.65ex\hbox
{$\sim$}}\;}
\def\gsim{\;\centeron{\raise.35ex\hbox{$>$}}{\lower.65ex\hbox
{$\sim$}}\;}

\def\super#1{\ifmmode \hbox{\textsuper{#1}}\else\textsuper{#1}\fi}
\def\textsuper#1{\newcount\holdspacefactor\holdspacefactor=\spacefactor
$^{#1}$\spacefactor=\holdspacefactor}

\def\getcite#1,{\advance\citenumber by1
\def\getcitearg{#1}\def\lastarg{@}
\ifnum\citenumber=1
\ref{#1}\let\next=\getcite\else\ifx\getcitearg\lastarg\let\next=\relax
\else ,\ref{#1}\let\next=\getcite\fi\fi\next}

\def\pom{{\rm P\kern -0.53em\llap I\,}}
\def\spom{{\rm P\kern -0.36em\llap \small I\,}}
\def\sspom{{\rm P\kern -0.33em\llap \footnotesize I\,}}

\relax
\newskip\humongous \humongous=0pt plus 1000pt minus 1000pt
\def\caja{\mathsurround=0pt}
\def\eqalign#1{\,\vcenter{\openup1\jot \caja
        \ialign{\strut \hfil$\displaystyle{##}$&$
        \displaystyle{{}##}$\hfil\crcr#1\crcr}}\,}
\newif\ifdtup

\def\eqright #1\cr{\noalign{\hfill$\displaystyle{{}#1}$}}
\def\eqleft #1\cr{\noalign{\noindent$\displaystyle{{}#1}$\hfill}}

\def\oldreffmt#1{\rlap{[#1]} \hbox to 2\parindent{}}

\def\figfmt#1{\rlap{Figure {#1}} \hbox to 1in{}}

\def\VEV#1{\left\langle #1\right\rangle}

\def\auto{\eqno(\refstepcounter{equation}\theequation)}
\def\begineq #1\endeq{$$ \refstepcounter{equation}\eqalign{#1}\eqno
        (\theequation) $$}
\def\contlimit{\,{\hbox{$\longrightarrow$}\kern-1.8em\lower1ex
\hbox{${\scriptstyle (a\rightarrow0)}$}}\,}
\def\centeron#1#2{{\setbox0=\hbox{#1}\setbox1=\hbox{#2}\ifdim
\wd1>\wd0\kern.5\wd1\kern-.5\wd0\fi
\copy0\kern-.5\wd0\kern-.5\wd1\copy1\ifdim\wd0>\wd1
\kern.5\wd0\kern-.5\wd1\fi}}
\def\centerover#1#2{\centeron{#1}{\setbox0=\hbox{#1}\setbox
1=\hbox{#2}\raise\ht0\hbox{\raise\dp1\hbox{\copy1}}}}
\def\centerunder#1#2{\centeron{#1}{\setbox0=\hbox{#1}\setbox
1=\hbox{#2}\lower\dp0\hbox{\lower\ht1\hbox{\copy1}}}}
\def\lsim{\;\centeron{\raise.35ex\hbox{$<$}}{\lower.65ex\hbox
{$\sim$}}\;}
\def\gsim{\;\centeron{\raise.35ex\hbox{$>$}}{\lower.65ex\hbox
{$\sim$}}\;}

\def\super#1{\ifmmode \hbox{\textsuper{#1}}\else\textsuper{#1}\fi}
\def\textsuper#1{\newcount\holdspacefactor\holdspacefactor=\spacefactor
$^{#1}$\spacefactor=\holdspacefactor}

\def\getcite#1,{\advance\citenumber by1
\ifnum\citenumber=1
\ref{#1}\let\next=\getcite\else\ifx#1@\let\next=\relax
\else ,\ref{#1}\let\next=\getcite\fi\fi\next}

\def\upon #1/#2 {{\textstyle{#1\over #2}}}
\relax

\def\subhead#1{\bigskip\vbox{\noindent\bf #1}\nobreak\par}

\def\til#1{\centeron{\hbox{$#1$}}{\lower 2ex\hbox{$\char'176$}}}
\def\tild#1{\centeron{\hbox{$\,#1$}}{\lower 2.5ex\hbox{$\char'176$}}}
\def\sumtil{\centeron{\hbox{$\displaystyle\sum$}}{\lower
-1.5ex\hbox{$\widetilde{\phantom{xx}}$}}}

\def\pom{{\rm P\kern -0.53em\llap I\,}}
\def\spom{{\rm P\kern -0.36em\llap \small I\,}}
\def\sspom{{\rm P\kern -0.33em\llap \footnotesize I\,}}

\begin{document}
\begin{titlepage}
\rightline{\vbox{\halign{&#\hfil\cr
&ANL-HEP-CP-94-46\cr
&\today\cr}}}
\vspace{0.25in}

\begin{center}

{\large \bf NEW STRONG INTERACTIONS ABOVE THE ELECTROWEAK SCALE}
\medskip

Alan R. White\footnote{Work
supported by the U.S. Department of
Energy, Division of High Energy Physics, Contract\newline W-31-109-ENG-38}
\\ \smallskip
High Energy Physics Division\\Argonne National
Laboratory\\Argonne, IL 60439\\ \end{center}

\begin{abstract}

      Theoretical arguments for a new higher-color quark sector, based on
Pomeron physics in QCD, are briefly described. The electroweak
symmetry-breaking, Strong CP conservation, and electroweak scale CP
violation, that is naturally produced by this sector is also outlined.

A further consequence is that above the electroweak scale there will be
a radical change in the strong interaction. Electroweak states, in
particular multiple $W$'s and $Z$'s, and new, semi-stable, very massive ,
baryons, will be commonly produced. The possible correlation of expected
phenomena with a wide range of observed Cosmic Ray effects at and above the
primary spectrum knee is described.

Related phenomena that might be seen in the highest energy hard scattering
events at the Fermilab Tevatron, some of which could be confused with top
production, are also briefly discussed.

\end{abstract}

\vspace{1.5in}

\noindent Invited talk presented at the 8th International Symposium on Very
High Energy Cosmic Ray Interactions, July 1994, Waseda University, Tokyo,
Japan.

\end{titlepage}

\subhead{\bf 1. INTRODUCTION}

The physics of very high energy Cosmic Rays, as seen in Mountain Emulsion
Chambers and Extensive Air Showers, is predominantly that of strong interaction
fragmentation and diffraction. It is well known in the Cosmic Ray community
that a significant number of effects now suggest the existence of new strong
interaction physics at energies around $10^{16}$ eV or higher. The most radical
proposal being\cite{nik,dy} that the famous ``knee'', in the induced incoming
energy spectrum, around this value is actually evidence for new physics rather
than a discontinous change in the primary spectrum. The corresponding center of
mass threshold for hadron-hadron scattering is $\sqrt{s} \sim 3-5$ TeV. This
implies that the new physical processes can probably not be seen directly at
the Fermilab Tevatron but, as I shall discuss, the physics involved might be
glimpsed in the highest energy virtual processes .

I have studied the QCD Pomeron responsible for diffraction for many years,
and for some time have advocated the theoretical necessity for a new, {\bf
higher-color}, quark sector. From my analysis\cite{arw1} this new sector is
directly required for the consistency of the Pomeron with both confinement and
perturbative QCD at high-energy. Remarkably such a sector can replace the
unaesthetic Higgs sector of the Standard Model (in partial analogy with
technicolor) and provide an essentially complete mechanism for mass
generation in the electroweak sector. This links the strong and electroweak
interactions in a direct manner and, in particular, implies that the
electroweak scale is explained as a second QCD scale. As I have studied this
possibility more and more seriously, I have gradually realized that other
deep puzzles of the Standard Model may also be resolved. For example, the
problems of Strong CP Conservation and CP Violation at the Electroweak
Scale.

In this talk I will first explain qualitatively why the new quark sector is
required in QCD and also describe the dynamical electroweak symmetry-breaking
and CP properties that result. I will then spend the remainder of the talk
elaborating on the essential feature for this conference. That is, not very
far {\bf above the electroweak scale there will be a radical change in the
strong interaction}. Electroweak states, in particular multiple $W's$ and
$Z's$, and new semi-stable baryons, will be commonly produced. We anticipate
that diffractive production of the new states will be a major (if not the
major) effect. Clearly this will dramatically change the nature of Cosmic
Ray showers and the states they produce above such energies. Although very
difficult to predict in any detail, I will suggest that the new phenomena
to be expected have the right characteristics to explain, qualitatively,
a wide range of observed Cosmic Ray effects, including the following.

\begin{itemize}

\item{Strong attenuation of family production, as observed in emulsion
chambers, together with a sharp change in the electromagnetic and hadronic
energy spectra.}

\item{Small $X_{max}$ for high-energy air showers with $E^0 \sim 10^{17}$ eV
together with a fast rise of $X_{max}$ as the energy increases.}

\item{Shorter ``hadronic'' interaction length in emulsion and lead chambers.}

\item{Anomalous penetration in the atmosphere and in detectors, often
involving the production of intense ``halos''.}

\item{Coplanarity of multi-halos.}

\item{Large $p_{\perp}$ production of ``Centauros'' - with low
electromagnetic energy, and ``Chirons'' - with apparent anomalously low
$p_{\perp}$ in secondary showers.}

\item{Excess of (underground) muon pairs with large separation.}

\item{Large zenith angle excess of high-energy air showers and azimuthal
asymmetry in $\gamma$ and hadron family production.}

\end{itemize}

In general the situation seems very interesting and a reasonable case can
be made that the type of modification of the strong interaction that I am
arguing for is actually being observed in the highest energy Cosmic Rays. Of
course, many of the above Cosmic Ray effects suffer from low statistics and it
will remain essential that they be observed in accelerator experiments if they
are to be confirmed and studied. The LHC will cover most of the relevant energy
range but it is a decade away from operating. As I discuss at the end of the
talk, it is also possible that the accumulating number of very high energy
hard scattering events at the Fermilab Tevatron (mostly involving photon and
weak vector boson states) could be a glimpse of the physics involved. Indeed
{\bf this physics might well be closely correlated with top production} and
there could be confusion, experimentally, between new and expected production
processes.

\vspace{.2in}

\subhead{\bf 2. POMERON FIELD THEORY}

In pre-QCD days the Pomeron was a phenomenological object - a Regge pole
which (essentially) was thought to reflect the low $k_{\perp}$ multiperipheral
production of multiple pions (with, say, $<n> ~ \sim 10-20$). As illustrated
in Fig.~1,
$$
\eqalign{
\sigma_T = \Sigma \int d\Omega_n |A_n|^2 ~\sim~ s^{\alpha_{\spom}(0) - 1}}
\auto
$$
so that $\sigma_T \sim C$ requires $\alpha_{\spom}(0) = 1$. {\bf Pomeron}
(reggeon) {\bf Field Theory} had both a phenomenological and, via reggeon
unitarity, a theoretical basis\cite{arw2} as an effective field theory
accounting for all the additional diffractive and absorptive effects that
unitarity requires must accompany multiperipheral pion production. High-mass
diffraction determines the magnitude of the triple Pomeron coupling.
Multi-Pomeron diagrams can be thought of as representing multiplicity
fluctuations. That is, an N-Pomeron state appearing on some part of the
rapidity axis is associated with a multiplicity fluctuation of N times the
average multiplicity in that rapidity interval. The origin of some reggeon
graphs is illustrated in Fig.~2.

The renormalization group can be applied to Pomeron Field Theory and, with a
triple Pomeron coupling, there is a {\bf Critical Pomeron} solution\cite{mpt}
for $\alpha_{\spom}(0) = 1$. This gives $\sigma_T ~\sim ~ (ln s)^{\eta}$ where
$\eta$ is an anomalous dimension. The Critical Pomeron is the only known
theoretical description of rising total cross-sections which satisfies {\bf
all $s$ and $t$ - channel unitarity constraints.}

Within Pomeron Field Theory we can study what happens if we initially violate
unitarity by setting $\alpha_{\spom}(0) > 1$. The result is a new {\bf
Super-Critical Pomeron} phase\cite{arw2} in which there is a Pomeron
condensate - giving rise to the vacuum production of Pomerons! At first
sight, it is difficult to understand how the vacuum production of Pomerons
could possibly have a physical interpretation. However, a detailed study of
the structure of the induced graphs,the transverse momentum singularities
that appear, and the resultant reggeon unitarity properties, leads to the
following remarkable result.

Vacuum production of Pomerons directly produces additional particle pole
factors in reggeon graphs that correspond to {\bf vector reggeon intermediate
states}. Consequently, the phase-transition when $\alpha_{\spom}(0) > 1$
involves the conversion of the divergences in rapidity (due to
$\alpha_{\spom}(0) > 1$) into particle divergences in transverse momentum. In
effect, if we try to make the total cross-section increase more rapidly than
allowed by unitarity, a {\bf vector particle V is ``deconfined''} and
appears in the theory coupling pair-wise to the Pomeron. The vector reggeon
trajectory $\alpha_V(t)$ is exchange-degenerate with the Pomeron trajectory
and, away from the critical point, the physical intercepts satisfy
$\alpha_{\spom}(0) = \alpha_V(0) < 1$. As a result, in both the Sub-Critical
and
Super-Critical phases, we have $\sigma_T \to~ 0$ asymptotically. This
implies that {\bf to obtain a rising cross-section the Pomeron must be
Critical}.

The Super-Critical Pomeron was discovered independently of QCD but a
fundamental question is, of course, what is the physical interpretation (if
any) of the Pomeron phase transition in QCD? Or, equivalently, can the vector
particle V be related to a (deconfined) gluon?

\vspace{.2in}

\subhead{\bf 3. THE POMERON PHASE-TRANSITION IN QCD}

In the multiperipheral model it is clear\cite{gt} that adding more hadron
states increases $\alpha_{\spom}(0)$. Therefore, {\bf if} there is a Regge
pole Pomeron in QCD, we anticipate that adding quarks moves the
Pomeron closer to criticality and that {\bf the Critical Pomeron might be
related to QCD with the ``maximum'' number of quarks.} All quarks become
(close to) massless in the asymptotic Regge limit, and so we are led to ask
- can the physics of QCD with a large number of massless quarks be related to
the Pomeron phase-transition?

Several properties of (massless) QCD suggest that the Critical Pomeron does
indeed occur when the number of flavors $N_f$ is a ``maximum''. Since these
can be described without introducing the technology of Reggeon Field Theory
we briefly describe them here. $N_f = N_f^{max} = 16$ is the maximum value
before asymptotic freedom is lost and (presumably) gluons are deconfined. (The
deconfinement implies, of course, that there is a phase transition but can we
identify the Critical Pomeron transition?) Consider first the behavior of the
$\beta$-function as a function of $N_f$. It is very likely\cite{bz,arw1} that
at $N_f = N_f^{max}$ (and probably only at this value) the $\beta$-function
develops an infra-red fixed-point, as illustrated in Fig.~3.

The scaling properties, and in particular, the variety of anomalous
dimensions that develop at this fixed-point clearly could be directly
related to the scaling properties of the Critical Pomeron. Indeed, it can be
shown\cite{arw1} that in the limit of zero meson mass (which we assume to
correspond to the limit of zero quark mass in QCD), the Critical Pomeron
forward diffraction peak behaves as illustrated in Fig.~4.
That is the diffraction pattern collapses into a simple peak that
has\cite{arw1} the character of a massless vector singularity with an
anomalous dimension, suggesting directly a relationship with a massless,
fixed-point, vector theory.

Also, when $N_f = N_f^{max}$ (and {\bf only} at this value), the first term
in the $\beta$-function is small enough\cite{gw} that adding a triplet
(``Higgs'') scalar to the theory retains asymptotic freedom of both
the gauge coupling and the scalar coupling. This implies that for this
``maximum'' number of quarks, the (dynamical?) Higgs mechanism can break the
gauge symmetry, from SU(3) to SU(2), without destroying the short-distance
properties of the theory. Consequently {\bf a vector reggeon V, i.e. a
massive gluon reggeon , can enter the theory smoothly} (with no $k_{\perp}$
cut-off in particular) {\bf only when when $N_f = N_f^{max}$}. Since the
entry of V into the theory characterises the Super-Critical Pomeron we have
another independent argument that $N_f = N_f^{max}$ is the critical
point.

Of course, we expect confinement to be crucial for the emergence of a
multiperipheral like Pomeron related to pion production etc.. The study of
confinement in the Regge limit is a major topic. After an elaborate
technical analysis utilising reggeon diagrams, it can be shown\cite{arw1}
that within QCD

\begin{itemize}

\item{\bf quarks play a crucial role in the simultaneous emergence of
confinement and a Regge pole Pomeron in the small $k_{\perp}$ high-energy
regime}

\item{\bf the energy-dependence of small $k_{\perp}$ physics (i.e.
$\alpha_{\spom}(0)$ )is strongly dependent on both $N_f$ and the $k_{\perp}$
cut-off and the Critical Pomeron occurs without a cut-off only when $N_f =
N_f^{max}$.}

\end{itemize}

To remove the $k_{\perp}$ cut-off and obtain a smooth matching with QCD
perturbation theory at large $k_{\perp}$, we must have a cross-section that
does
not go to zero asymptotically (i.e. $\alpha_{\spom}(0) = 1$), since this is
what
the perturbation expansion gives. Therefore, for confinement and QCD
perturbation theory to coexist in the high-energy region, we must obtain the
Critical Pomeron as the large $k_{\perp}$ cut-off is removed. Since this
requires $N_f = N_f^{max}$, {\bf a further quark sector must exist!!}

It is very important, however, that (assuming six flavors are known to
exist)

\begin{itemize}

\item{\bf the further quark sector need not be 10 more flavors of color triplet
quarks. $N_f = N_f^{max}$ is also produced by an additional flavor
doublet of color sextet quarks.}

\end{itemize}

\vspace{.2in}

\subhead{\bf 4. ELECTROWEAK SYMMETRY BREAKING, ELECTROWEAK SCALE INSTANTON
INTERACTIONS, AND CP-VIOLATION}

{}From the perspective of QCD Pomeron physics, it is a remarkable coincidence
that two flavors of color sextet quarks can provide\cite{wjm} a natural form
of dynamical symmetry-breaking for the electroweak interaction which meshes
perfectly with the observed experimental features. Indeed this provides a
self-contained motivation for introducing the higher color quark sector
which we can briefly outline as follows.

Consider adding to the Standard Model (with no scalar Higgs sector), a massless
flavor doublet $(u_6,d_6)$ of color sextet quarks with the usual quark
quantum numbers, except that the role of quarks and antiquarks is
interchanged. For the $SU(2)\otimes U(1)$ anomaly to be cancelled there must
also be other fermions with electroweak quantum numbers added to the
theory\cite{wjm,kk}, but we shall not consider this here except to note that
this could be the color octet leptoquarks discussed below. We consider first
the
QCD interaction of the massless sextet quark sector. There is a $U(2)\otimes
U(2)$ chiral flavor symmetry. QCD chiral dynamics will break the axial
symmetries spontaneously and produce four massless pseudoscalar mesons
(Goldstone bosons), which we denote as $\pi^+_6,\;\pi^-_6,\;\pi^0_6$ and
$\eta_6$, in analogy with the usual notation for mesons composed of $u$ and
$d$ color triplet quarks.

As long as all quarks are massless, QCD is necessarily $CP$ conserving in
both the sextet and triplet quark sectors. Therefore, in the massless
theory we can, in analogy with the familiar treatment of flavor isospin
in the triplet quark sector, define sextet quark vector and axial-vector
currents $V^{\tau}_{\mu}$ and $A^{\tau}_{\mu}$ which are ``isotriplets'' under
the unbroken $SU(2)$ vector flavor symmetry and singlet currents
$v_{\mu}$, $a_{\mu}$. The pseudoscalar mesons couple ``longitudinally'' to the
axial currents, that is
$$
\eqalign{ <0|A^\tau_{\mu}|\pi^{\tau}_6(q)>~\sim F_{\pi_6}q_{\mu}~~,
{}~~~~~~<0|a_{\mu}|\eta_6(q)>~\sim F_{\eta_6}q_{\mu}}
\auto
$$
while the vector currents remain conserved. (Note that $a_{\mu}$ should
actually
contain a small admixture of the triplet quark flavor singlet axial current if
it is to generate the U(1) symmetry orthogonal to that broken by the QCD
$U(1)$ anomaly).

We consider next the coupling of the electroweak gauge fields to the sextet
quark sector. The massless $SU(2)$ gauge fields $W^{\tau}_{\mu}$ couple to the
isotriplet sextet quark currents in the standard manner, that is
$$
\eqalign{{\cal L}_I=gW^{\tau\mu}\Bigl(V^{\tau}_{\mu}-A^{\tau}_{\mu}\Bigr)}
\auto
$$
It follows from (2) and (3) that the $\pi^+_6,\;\pi^-_6$ and $\pi^0_6$ are
``eaten'' by the $SU(2)$ gauge bosons and (after the hypercharge interaction is
included) respectively become the third components of the $W^+,\;W^-$ and
$Z^0$. Consequently, QCD chiral symmetry breaking generates masses for the
$W^+,\;W^-$ and $Z^0$ with $M_W\sim g\;F_{\pi_6}$ where $F_{\pi_6}$ is {\em a
QCD scale}. We anticipate that the relative scales of triplet and sextet
chiral symmetry breaking are determined by the ``Casimir Scaling''
rule\cite{wjm}, i.e. if $C_6$ and $C_3$ are sextet and triplet Casimirs
respectively, then
$$
\eqalign{C_6\alpha_s(F^2_{\pi_6})~\sim~C_3\alpha_s(F^2_{\pi})}
\auto
$$
which is consistent with $F_{\pi_6}\sim 250$ GeV!

We conclude that a sextet sector of $QCD$ produces a special version of
``technicolor'' symmetry breaking in which {\bf the electroweak scale is
naturally explained as a second $QCD$ scale}. Also since we are completely
restricted to a flavor doublet the form of the symmetry-breaking is
automatically equivalent to that of an $SU(2)$ Higgs sector and so
$$
\eqalign{\rho=~(M^2_W/M^2_Zcos^2\theta_W)~=~1}
\auto
$$
as required by experiment.

Therefore introducing a sextet quark sector not only produces a matching of
the asymptotic freedom and confinement properties of $QCD$ via the Critical
Pomeron, but also gives a natural solution to the major problem of today's
Standard Model i.e. the nature of electroweak symmetry breaking. {\bf The
sextet
sector may}, as we now discuss, {\bf also be deeply tied up\cite{kkw,arw3}
with the issue of Strong $CP$ conservation.}

The $\eta_6$ is not involved in generating mass for the electroweak gauge
bosons, but instead remains as a Goldstone boson associated with a $U(1)$
axial chiral symmetry. It is therefore an axion\cite{wil} in the original sense
of the Peccei-Quinn mechanism\cite{pq} and it remains massless until triplet
quark masses are added to the theory. In the present context, this involves
the addition of triplet/sextet four-fermion couplings (that should ultimately
be traceable to a larger unifying gauge group), which, when combined with
the sextet quark condensate, provide triplet quark masses. That $CP$ remains
conserved by QCD triplet quark interactions then follows from the original
Peccei-Quinn argument utilising the sextet axial $U(1)$ symmetry.

At this stage another very important property of QCD with $N_f = N_f^{max}$
is crucial. It seems that renormalon singularities are completely absent in
the Borel plane\cite{arw1,cjm}. This implies that perturbation theory is
much more convergent and that {\bf instanton interactions are both infra-red
finite and provide all the non-perturbative physics} of the theory.
Instanton interactions are therefore well-defined at the lowest infra-red
scale of the theory, i.e. the electroweak scale. Combining this with the
extremely slow evolution of the gauge coupling, the instanton interactions
are then enhanced by integration over an extremely wide size range  (for the
instanton involved). Consequently the $\eta_6$ can aquire a large, i.e.
electroweak scale, mass as a result of electroweak scale (and  higher) color
instanton interactions\cite{arw3} and, unlike a conventional Peccei-Quinn
axion, is certainly not ruled out experimentally. Indeed it may even have
been seen\cite{kkw}. Clearly all of the particular properties of QCD with
color sextet quarks play an intrinsic role in this very special resolution
of the Strong CP problem.

A rather complicated set of fermion vertices is actually generated
by the electroweak scale instanton interactions. Because of the distinct
Casimirs involved, the singlet current
$$
\eqalign{J^0_{\mu}~=~a^6_{\mu}-5a^3_{\mu}}
\auto
$$
is conserved in the presence of instantons (6 and 3 now denote sextet and
triplet currents respectively). Consequently the minimum instanton
interaction involves one quark/antiquark pair of each triplet flavor and
five pairs of each sextet flavor. Combining this interaction with the existence
of both sextet and triplet chiral condensates (and, also, four-fermion
vertices coupling triplets and sextets) a wide assortment of fermion
vertices is produced. As we discuss further in the next Section, we expect
that these vertices will play a major role in strong interactions above the
electroweak scale.

Finally we note that {\bf the sextet sector may also be responsible for
$CP$ violation at the weak scale}. Because the {\bf sextet sector has no
axion} the QCD interactions at this scale will naturally be {\bf ``Strong
$CP$-violating''. The familiar triplet quark hadrons will contain a small
admixture of sextet quark states - which could provide their $CP$ violating
interactions}.

Before we go on to the the new strong interactions and their consequences for
Cosmic Ray physics, we would like to emphasize the (unconventional) implication
of the foregoing arguments. Namely that understanding the intricacies of the
strong interaction may actually provide answers to remaining problems of the
weaker interactions. Or equivalently

\begin{itemize}

\item{\bf the QCD Pomeron may be the Key to Many of the
Remaining Puzzles of the Standard Model}

\end{itemize}

\vspace{.2in}

\subhead{\bf 5. THE NEW STRONG INTERACTIONS}

Above the sextet chiral scale, that is the electroweak scale, the sextet
sector will be a major part of the QCD interaction. $QCD_{max}$ (that is QCD
with $N_f = N_f^{max}$ - via the triplet and sextet sectors) is a very
different gauge theory to those conventionally studied. The gauge coupling
is relatively small and effectively does not run. While the sextet sector
can, presumably, be integrated out to give conventional QCD at low energies,
at high energy we can expect very different behavior. Some ``non-perturbative''
physics will perhaps be understood via conventional non-perturbative QCD
ideas in terms of sextet flux tubes etc.. However, many non-perturbative
effects will surely be directly dependent on the multitude of higher-order
(instanton) fermion interactions involving sextet quarks. (We have emphasized
that these interactions are enhanced by the gauge coupling not running). As
illustrated in Fig.~5, these interactions will generate high-order vertices
coupling $W$'s, $Z$'s, and $\eta_6$'s, with
$$
{\openup3\jot
\eqalign{ \Gamma_{mW,~nZ,~r\eta_6}~~
&\sim ~~~{F_{\pi_6}~(momentum scale)^2 \over \VEV{q_6\bar{q}_6}} ~~~
\Gamma_{(m-1)W,~nZ,~r\eta_6}\cr
&\sim ~~~{F_{\pi_6}~(momentum scale)^2 \over \VEV{q_6\bar{q}_6}} ~~~
\Gamma_{mW,~(n-1)Z,~r\eta_6}\cr
&\sim ~~~{F_{\pi_6}~(momentum scale)^2 \over \VEV{q_6\bar{q}_6}} ~~~
\Gamma_{mW,~nZ,~(r-1)\eta_6}\cr}}
\auto
$$
where $\VEV{q_6\bar{q}_6}$ is the sextet condensate and
$(\VEV{q_6\bar{q}_6})^{{1 \over 3}}~\sim~ F_{\pi_6}~\sim$ 250 GeV.

The $W^{+,-}$, $Z^0$ and $\eta_6$ are the ``PIONS'' of the sextet sector
and, as we have just described, they will be multiply produced, via a
``hard'' interaction at the electroweak scale. Since the mass and decay
properties of the $\eta_6$ are not well understood\cite{kkw} and it has, of
course, not yet been discovered, we will concentrate mainly on multiple $W$
and $Z$ production.

Because of the Casimir effect, we anticipate that sextet states will have a
stronger coupling to gluons, and hence to the Pomeron, than does the triplet
sector. Therefore

\begin{itemize}

\item{\bf sextet states will have larger hadronic cross-sections than triplet
states (i.e. conventional hadrons)}

\end{itemize}

My work\cite{arw1} on high-energy hadrons interacting via the Pomeron can be
heuristically understood if we visualize a hadron as a conventional bag
containing quarks but with the surface containing a ``topological
condensate'' due to instanton interactions and expanding as illustrated in
Fig.~6.

\vspace{4in}

\begin{itemize}

\item[{Fig.~6}] Heuristic picture of a high-energy hadron and the Pomeron
as perturbative gluon exchange in a topological condensate background.

\end{itemize}

In first approximation, the Pomeron can then, as illustrated, be
thought of as one gluon exchange within the overlapping topological gauge
fields of the scattering hadrons. (Note that this automatically gives the
``additive quark model'' result that the Pomeron couples directly to a
single quark in a hadron).

The topological gauge fields of the hadrons will also, via instanton
interactions, be responsible for multiple $W$ and $Z$ production
accompanying the perturbative gluon interaction. Therefore we
conclude that a major component of the new strong interactions above the
electroweak scale will be

\begin{itemize}

\item{\bf diffractive production, with very high transverse momentum, of states
containing large numbers of $W$'s and $Z$'s.}

\end{itemize}

\noindent This will be one major ingredient of our discussion of Cosmic Ray
effects.

Next we note that the sextet quark sector will produce new BARYONS
of the form
$$
\eqalign{~\bar{q}_6qq~,~q_6\bar{q}\bar{q}~,~\bar{q}_6\bar{q}_6q~,
{}~q_6q_6\bar{q}~,~\bar{q}_6\bar{q}_6\bar{q}_6~,~
q_6q_6q_6~,}
\auto
$$
There will also be VECTOR MESONS of the form
$$
\eqalign{q_6\bar{q}_6}
\auto
$$

The sextet quark constituent mass is presumably of the same order of
magnitude or a little larger than the chiral scale and so for definiteness
we will take it to be $\sim 400$ GeV. Clearly the lightest new states will
be the BARYONS containing just one sextet quark. Their mass will be very
close to the sextet mass i.e. $\sim 400$ GeV and since, within $QCD_{max}$,
sextet and triplet baryon numbers are separately conserved, they will be
very stable. We refer to BARYONS containing two (triplet) quarks as $P$'s and
those containing two antiquarks as $\bar{P}$'s. The VECTOR MESONS will decay
into the PIONS of the theory and so will give resonance production of $W$'s,
$Z$'s and $\eta_6$'s at the TeV scale. The higher mass BARYONS will
presumably decay into $P$'s and $\bar{P}$'s (together with appropriate
combinations of normal hadrons).

For the next Section it will be crucial that {\bf the $P$'s and $\bar{P}$'s
are sufficiently stable that sometimes (but not always) they survive a trip
(with collisions) from near the top of the atmosphere down to mountain-top
detectors.}

If the $P$'s and $\bar{P}$'s are to decay, there must be a further (unifying)
interaction coupling the two distinct quark sectors. At first sight this
could be a high mass (GUT) gauge boson. But the absence of proton decay
probably makes it very difficult to construct such a theory consistently if
the BARYONS are to decay much faster than protons! An alternative\cite{kk}
is that within the unified theory, there are further color octet quarks
($q_8$) (these could be ``leptoquarks'') that enter at a mass scale just a
few orders of magnitude above the electroweak scale. At this scale the
unified theory can be asymptotically free even though the QCD subsector
will not be. If the unified theory is chiral then in general two-fermion
condensates are not gauge invariant. Gauge-invariant condensates must
contain at least four fermion fields and so it is natural\cite{kk} to expect
that, at the new high mass scale, condensates of the form
$$
\eqalign{ \VEV{q_8q_8q_6q_3}}
\auto
$$
will exist. QCD instanton interactions, at this scale, will then produce the
appropriate sextet quark decays. We therefore assume that

\begin{itemize}

\item{\bf $P$'s and $\bar{P}$'s are ``semistable'' with a decay rate
determined by a mass scale much larger than the electroweak scale. We
anticipate
that their decay modes will include states containing multiple $W$'s and
$Z$'s.}

\end{itemize}

Clearly BARYONS can be pair-produced diffractively by the
Pomeron (in particular, via an instanton interaction), also VECTOR MESONS
can be produced with an accompanying $W$'s and/or $Z$'s. From the experimental
evidence on the diffractive production of strange baryons\cite{R608}
illustrated in Fig.~7, we can assume that this production process will have
some important properties, which we can explain as follows.

Because the Pomeron couples predominantly to a single quark in a proton, two
constituent quarks persist in the forward direction of the initial proton
(with around 90\% probability) during any diffractive excitation process. If
a new forward going baryon is to be formed then this is achieved by the
vacuum production of additional quark-antiquark pairs in the center of mass
of the scattered quark and the forward going diquark system. (This process
can be an instanton interaction). There are two consequences of this
production mechanism which will carry over directly into the diffractive
production of BARYONS.

Firstly, because only a single quark can be replaced in the fast proton
if a BARYON-ANTIBARYON pair is produced
\begin{itemize}

\item{\bf there is a charge bias in the production of the forward produced
BARYON - it is necessarily positively charged or neutral. Correspondingly
the charge of the ANTIBARYON state produced away from the forward direction is
either negative or neutral.}

\end{itemize}

Secondly, if all vacuum pairs involved are produced (almost) at rest in
the center of mass of the scattered quark and diquark system then, as is
illustrated by the data for diffractive production of
$\Lambda^0\bar{\Lambda}^0$ pairs shown in Fig.~7,

\begin{itemize}

\item{\bf the full diffractively produced state is approximately coplanar - it
lies in the plane formed by the momenta of the forward going fast BARYON and
the (ANTI-)BARYON with the smallest forward momentum.}

\end{itemize}

With all of the properties highlighted in this Section in hand we can now
try to explain at least part of the wide range of Cosmic Ray exotica.

\vspace{.2in}

\subhead{\bf 6. COSMIC RAY EVIDENCE FOR THE NEW STRONG INTERACTIONS.}

In this Section I will go through the phenomena I listed in the
Introduction, giving a brief summary of the experimental results and then
describing their interpretation in terms of the physics of the last Section.
Since I am not an expert I may well have misunderstood some of the phenomena
involved. If so I apologize to the authors involved.

\begin{center}

{\bf Strong attenuation of family production, as observed in emulsion
chambers, together with a sharp change in the electromagnetic and hadronic
energy spectra.}

\end{center}

Such effects have been seen by the Chacaltaya and Pamir collaborations and
more distinctively in the highest energy results of the HADRON experiment at
Tien-Shan. Figs.~8(a) and 8(b) show that the combination of low family flux
and small energy spectrum indices for constituent showers in the
Chacaltaya/Pamir data\cite{cp} is not fit by any of the conventional models.
As shown in Fig.~8(c), the discrepancy is less if a heavy nuclei primary
composition is assumed.

Fig.~9(a) shows a possibly related effect in the data\cite{nik} from the HADRON
experiment. The $\gamma$ spectrum of shower cores scales up to a certain energy
and then softens as the energy increases. As is shown, the softening could be
reproduced by a heavy primary composition but the
overall intensity would be much too high. Fig.~9(b) shows that the
assumption of a heavy primary composition is inconsistent with the muon
multiplicity distribution obtained at Tien-Shan.

The change of the $E_{\gamma}$ spectrum suggests the existence of a physical
threshold around the knee energy. It also implies that, for some fraction of
the Chacaltaya/Pamir events, the primary energy may be higher than given by
conventional physics models. Therefore new physics above the threshold may
be involved in these events also.

My explanation of these phenomena is close to that already suggested by
those working on the HADRON experiment\cite{nik}. At high enough energies,
production of the heavy, semistable, $P$'s and $\bar{P}$'s will be a
significant part of the diffractive and fragmentation cross-sections. Since
these BARYONS are semi-stable they will propagate for large distances within
the shower, sometimes reaching the detector. The evolution of that part of
the shower energy not in the heavy BARYONS wil be normal but will clearly
produce far fewer gamma and hadron families. This effect is labeled
``fragmentation region disappearance" by Nikolsky\cite{nik} who argues that
the particles involved should have a mass $\geq$ 400 GeV. This explanation
achieves the same effective reduction of primary energy as heavy nuclei
primaries would do, but without the high multiplicity muon production that
is not seen in Fig.~9(b).

A heavy primary composition for energies around $10^{16}$ GeV is also
incompatible with the Soudan 1 underground muon multiplicity\cite{soud}.
Recent MACRO data\cite{mac} shown in Fig.~10(a) leads to a similar
conclusion. As is shown in Fig.~10(b), the high multiplicity tail of this
distribution is determined by the highest primary energies, whatever
(conventional physics) composition model is utilised. {\bf The absence of a
large number of high multiplicity muon events is clearly a major problem for
any explanation of very high energy Cosmic Ray phenomena that apppeals to a
large heavy nuclei composition.}

\begin{center}

{\bf Small $X_{max}$ for high-energy air showers with $E^0 \sim 10^{17}$ eV
together with a fast rise of $X_{max}$ as the energy increases.}

\end{center}

This is the Fly's Eye result\cite{fly}. As illustrated in Fig.~11, results
for the lowest energies i.e. $E^0 \sim 10^{17}$ eV, give a sufficiently low
average value for $X_{max}$ that a very strong heavy nuclei composition has
to be used to fit the data with conventional physics models. However, as the
energy rises this average increases too fast and the distribution changes
too much for a single composition model to fit the data. It is necessary to
vary the composition with energy as illustrated. The initial heavy nuclei
composition is again at variance with the lack of high-multiplicity
underground muon events mentioned above.

My explanation here is, in part, the same as for the previous effect. At the
lower end of the energy range the production of the heavy, semistable, BARYONS
will reduce the development of the shower and the consequent average $X_{max}$
in the same manner as the heavy nuclei composition. However, as we get to
energies high compared even to the sextet scale we can expect that, in analogy
with the triplet sector, high multiplicity PION states will be the dominant
sextet states produced. That is the production of $W$'s, $Z$'s and
$\eta_6$'s will dominate. Since these states are all unstable the showers
will develop more like normal proton showers. This could produce naturally
the required energy dependence of the $X_{max}$ distribution without any
dramatic change in composition.

\begin{center}

{\bf Shorter ``hadronic'' interaction length in emulsion and lead
chambers.}

\end{center}

The results of the Chacaltaya/Pamir collaboration\cite{cp1} are shown in
Fig.~12. Both in the Chacaltaya emulsion chambers and in the Pamir lead
chambers there is a pronounced decrease in the hadronic interaction length in
the highest energy showers.

I attribute this in part to the higher hadronic cross-section of those sextet
BARYONS that reach the detector. Also multiple $W$, $Z$ and $\eta_6$
intermediate states will produce major decay modes involving heavy flavors,
leptons, and photons, which may be partly responsible for the effect.

\begin{center}

{\bf Anomalous penetration in the atmosphere and in detectors,
involving the production of intense ``halos'' in the highest energy showers.}

\end{center}

Examples of events which have extreme penetration in lead\cite{cp1} and in
emulsion chambers\cite{cp}  are shown in Fig.~13. Some of them continue
producing new showers down to an extraordinary depth. At the highest
energies the cores of such showers contain very intense halos recorded on
X-ray films.

These effects have to be produced by BARYONS that enter the detector.
Multi-halo events should presumably be interpreted as involving
multi-BARYON states, although the initial production of very energetic
$W$'s and $Z$'s could also be involved. In the very highest energy events
there could even be VECTOR MESON resonances.

\begin{center}

{\bf Coplanarity of multi-halos.}

\end{center}

The coplanarity of multi-halos in very high energy showers is a striking
phenomenon having an established statistical significance. Results from the
Pamir collaboration\cite{pam} are illustrated in Fig.~14. The X-rays for
individual events are shown as well as the energy-dependence of the
alignment. A table also illustrates how conventional models fail to produce
the alignment. The experimenters emphasize that the total cross-section for
halo events is far too large for them to originate from minijet
configurations\cite{hal}.

My description of the alignment phenomenon in diffractive production of
BARYONS provides a direct explanation of this phenomenon. It is the same as
is seen in the diffractive production of strange baryons at the ISR!

\begin{center}

{\bf Large $p_{\perp}$ production of ``Centauros'' - with low
electromagnetic energy, and ``Chirons'' - with apparent anomalously low
$p_{\perp}$ in secondary showers.}

\end{center}

Familiar plots of hadronic versus electromagnetic energy for
Chacaltaya/Pamir\cite{cp} data and the comparison with simulations are shown in
Fig.~15. Centauro events represent the extreme of a general phenomenon that
less electromagnetic energy is produced than in normal pion production
events. The overall $p_{\perp}$ involved is apparently large but from their
narrowness, the $p_{\perp}$ in secondary showers appears to be anomalously
small. The general class of events with these $p_{\perp}$ properties are
referred to as Chirons.

As we have said, BARYONS will generally be produced in the initial
atmospheric collision of the Cosmic Ray primary. We can assume they will
sometimes decay directly just above or in the detector. Often they will
undergo secondary collisions and then decay similarly. The collisions and
(or) the decay will involve very high initial $p_{\perp}$ and can take place
sufficiently close to the detector that secondary $p_{\perp}$ within
produced showers is normal even though they appear anomalously narrow. Since
multiple $W$, $Z$ and $\eta_6$ intermediate
states will again be involved, we can anticipate that in general the
production of heavy flavors, taus and muons, will produce final states that
will be interpreted as anomalously ``hadron-rich'' in the detector.

\begin{center}

{\bf Excess of (underground) muon pairs with large separation.}

\end{center}

The underground muon experiments also measure the distribution of the distance
separation of muon pairs. The MACRO distributions\cite{mac1} are shown in
Fig.~16. There is an apparent  excess at large distances which would not be
expected from conventional physics models. Not surprisingly, I would like to
interpret the excess as evidence that $Z^0$'s are being directly produced,
with a hadronic cross-section, in high-energy Cosmic Rays. Potentially $Z^0$
events could be explicitly identified. This could provide, strong, direct
evidence for new physics such as I am proposing.

\begin{center}

{\bf Large zenith angle excess of high-energy air showers and azimuthal
asymmetry in $\gamma$ and hadron family production.}

\end{center}

Finally I come to some further results from the HADRON experiment. Fig.~17
shows\cite{dy} the zenith angle dependence for high-energy showers at two
energies. The straight lines are
conventional physics simulations at the two energies. There is a clear
excess at large zenith angle at the highest energy. Also shown, in
Fig.~17(b), is the zenith angle dependence of a break in the general size
(energy) spectra of the showers. It is interesting that the break is
essentially independent of the zenith angle and is located at the energy of
the knee, in the conventionally induced primary energy spectrum. Since
showers at different angles degrade differently in the atmosphere this, in
itself, suggests that there is some physics effect in the break which is not
simply related to the primary spectrum. Indeed it clearly leads to the
suggestion\cite{nik,dy}, referred to in the Introduction, that there is a
``new physics'' effect involved in the knee, and not just a simple change in
spectrum.

In Fig.~18 we show the most exotic (and, if it should be confirmed, perhaps the
most exciting) result from the HADRON experiment. A striking asymmetry in
the azimuthal angular dependence\cite{che} of the large zenith angle showers
is shown. This could perhaps be explained\cite{che} as due to the earth's
magnetic field if massive, {\bf negatively charged}, particles are
preferentially responsible for the showers.

I interpret these last results as evidence that in initial atmospheric
collisions secondary, semi-stable, very energetic, particles are produced
at varying (relatively) large angle, which are then interpreted as
(separated) large zenith angle primary showers. If such secondary particles
are responsible for the excess, this could explain why the spectrum break
is at the same shower size independently of the zenith angle. My proposal
is, of course, that the secondary particles are BARYONS. According to our
diffractive production argument above, the larger angle BARYONS contributing
to the larger zenith angle showers will be {\bf preferentially negatively
charged.}

It seems possible therefore that {\bf the azimuthal asymmetry could be
explained
by the charge asymmetry of the larger angle versus forward angle diffractive
production of BARYON pairs} described in the last Section.

\vspace{.2in}

\subhead{\bf 7. NEW HARD SCATTERING PROCESSES AND TOP PRODUCTION.}

If the new physics seen in Cosmic Rays were simply diffractive production of
$W$ and $Z$ pairs (or perhaps $\eta_6$ pairs) then we might estimate the
effective threshold to be, say, $x = (1 - M^2/s) \geq 0.96$ i.e.
$$
\eqalign{ \sqrt{s}~~\geq~~5~M ~~&\sim~~5~\times~160~~ GeV\cr
&\sim ~~800~ GeV}
$$
and so to be visible at the Fermilab Tevatron. Indeed, it remains possible that
$W$ pairs are produced, in some number, relatively far forward since this would
be impossible to determine with the present detectors.

{}From our discussion in previous sections it is clear that the more
distinctive
effects involve at least BARYON pair production. The corresponding diffractive
threshold would then be roughly
$$
\eqalign{ \sqrt{s}~~\geq~~5~M ~~&\sim~~5~\times~800~~ GeV\cr
&\sim ~~4~ TeV}
$$
which is consistent with the Cosmic Ray effects. Of course, we can also
expect some effects of the new sector to show up at energies well below the
diffractive threshold.

Indeed we might expect the new quark sector to first show up in the highest
transverse energy (but very rare) hard scattering events. Instanton
interactions will provide transitions from the (light) triplet quark sector,
to the sextet sector. Amongst the simplest possible states that can be
produced are $W^+W^-$, $Z^0Z^0$, $\eta_6\eta_6$, $Z^0\gamma$ and
$\gamma\gamma$. There were indications from UA1\cite{UA1} that the hard
scattering cross-section for $W$ pairs is indeed anomalously large. The
events at CERN were detected in the $WW \to $ leptons + 2 jet channel,
which, of course, has a relatively large branching ratio. However, at the
Tevatron this channel may be obscured since the background from conventional
QCD processes is much larger than at the CERN collider. Nevertheless an
excess of very high energy hard scattering events may be accumulating at the
Tevatron (including $Z^0\gamma$ and $\gamma\gamma$ events). Although these
events are not yet statistically significant, they may become so as the
experiments continue to take data.

Enhanced electroweak scale instanton interactions can provide transitions
from the familiar (light) triplet quark sector, not only to the sextet
sector we have been discussing, but also to states that are a combination of
sextet and (preferentially) heavy triplet quarks, and even to purely triplet
heavy quark states. This implies that the top quark (with a mass of $\sim$
170 Gev according to recent CDF results\cite{cdf}) may have a larger production
cross-section than standard perturbative estimates would give. Additional
states that can be produced include
$$
\eqalign{ W^+W^-~+ ~b\bar{b}~,~~~~Z^0Z^0~+~b\bar{b}~, ~~~...}
$$
The first state can clearly be directly confused with top production.
Indeed in the CDF analysis\cite{cdf} searching for candidate top events, a
few events have been found which are candidates to be identified with the
second final state. In many respects, these events strongly resemble those
identified as top events, but they should not be present at all according to
the Standard Model. This clearly suggests that some of the candidate top
events might in fact be direct $WW + b\bar{b}$ events. CDF also has a clear
$WZ$
event which has a very low probability to occur, according to the Standard
Model. An instanton interaction has to conserve charge in the sextet sector
but could produce a $WWZ$ state, with one $W$ in a region of phase space where
it escapes detection.

We conclude that {\bf a glimpse of sextet quark physics at the Tevatron
collider
may have already been provided}. As data is accumulated it should become
clear whether this is indeed the case. If it is, {\bf there will be a lot
more than top quark production that provides ``new physics'' in the
highest-energy hard scattering events.}

\newpage

\noindent{\bf Figure Captions}

\begin{itemize}

\item[{Fig.~1}] Multiperipheral pion production generating the Regge pole
Pomeron.

\item[{Fig.~2}] Varying multiplicity densities on the rapidity axis generate
higher-order Pomeron diagrams.

\item[{Fig.~3}] Evolution of the $\beta$-function with $N_f$ - (a)
$N_f \sim 5,6$ (b) $N_f \sim 14,15$ (c) $N_f = 16$

\item[{Fig.~4}] The Critical Pomeron asymptotic diffraction peak appproaches
a simple peak if quark masses are sent to zero.

\item[{Fig.~5}] Instanton interactions combined with condensates generate
multiple vertices for $W$'s, $Z$'s and $\eta_6$'s,

\item[{Fig.~7}] Coplanar diffractive production of strange baryons in
experiment R608 at the ISR.

\item[{Fig.~8}] Relation between the the family flux and power indices of
the energy spectrum $\beta$ (a) for single core showers and (b) for shower
clusters, in the energy range of 10-50 TeV, for the joint (J), Pamir (P) and
Chacaltaya (C) chambers, compared to simulation models. (b) The single-core
comparison when a heavy primary composition is used in the models.

\item[{Fig~9}] (a) Energy spectra of $\gamma$-quanta and electrons in EAS
cores with varying primary energies. The wide shaded strip is a simulation
with primary protons, the narrow strip is with a heavy nuclei composition.
(b) The experimental muon multiplicity distribution compared to simulations
with 1) heavy nuclei primary composition 2) an increasing inelasticity
coefficient and 3) the energy of the fragmentation region is lost from the
hadron-electron cascades.

\item[{Fig.~10}] (a) The underground muon multiplicity distribution results
from the MACRO experiment. (b) The relationship between muon multiplicities
and primary energy in conventional physics simulation models.

\item[{Fig.~11}] Average $X_{max}$ as a function of primary energy. Black
dots : data. Open squares : simulation with a proton dominant primary
composition. Open circles : simulation with a dominant heavy nuclei
composition. Diamonds : simulation with an energy-dependent composition.

\item[{Fig.~12}] (a) Distribution of shower starting position for
Chiron-type families observed by Chacaltaya two-storey chambers. The dotted
line represents exponential decrease with the geometrical attenuation mean
free path. (b) The $\Delta$T distribution of 170 hadrons with
$E_{\gamma} \geq 10$ TeV in 16 high energy famlies with
$E_{family} \geq 700$ TeV observed in the thick lead chambers of the Pamir
experiment. (c) For comparison the $\Delta$T distribution of all hadrons in
the Pamir lead chamber.

\item[{Fig.~13}] (a) Shower transitions with spot darkness plotted against
depth for two events in the Pamir lead chambers (b) Examples of strongly
penetrating, small spread, shower clusters in Chacaltaya chambers.

\item[{Fig.~14}] a) Darkness contours on X-ray films for 6 Pamir multicore
halo events. b) The energy dependence of the percentage of four-core events
satisfying $\lambda_4 \geq 0.8$ where $\lambda_4$ is a suitably defined
alignment parameter. c) Table comparing alignment percentages for
simulations and experimental data.

\item[{Fig.~15}] (a) The upper graph is a scatter plot of the number of
hadrons, $N_h (E_h^{(\gamma)} \geq 4$ TeV) in a family and the fraction of
the family energy carried by hadrons $Q_h ~(\equiv \sum E_h^{(\gamma)} /
\sum E_{(\gamma)} + E_h^{(\gamma)})$. Closed circles are for 173 families
in the joint chambers and 135 families in the Pamir chambers, and open
circles are for 121 families in the Chacaltaya chambers. The lower graph is
a simulation. The primary composition assumed is - solid black dot = proton,
open dot = alpha, diamond = CNO, x = heavy, + = iron. (b) Is the same as (a)
but with the families selected to have lateral spread
$\VEV{E^*R^*} < 300$ GeV.m.

\item[{Fig.~16}] The MACRO distance separation for muon pairs compared
with simulations (a) for muon pairs within all muon events (b) for dimuon
events only.

\end{itemize}

\end{document}